\documentclass[preprint,12pt]{elsarticle}

\usepackage{amssymb}
\usepackage{amsfonts}
\usepackage{amsmath}
\usepackage{graphicx}
\usepackage{mathrsfs}
\usepackage[hypertex,linkcolor=red]{hyperref}
\def\<{\langle}\def\d{\operatorname{d}}
\def\>{\rangle}\def\map#1{\mathscr{#1}} 
\def\SU2{\mathbb{SU}(2)}
\def\sH{\mathcal{H}}
\def\Tr{\operatorname{Tr}}
\def\d{\operatorname{d}}

\journal{}

\begin{document}

\begin{frontmatter}


  \title{Quantum No-Stretching: a geometrical interpretation of
    the No-Cloning Theorem}


\author[dip,infn]{Giacomo M. D'Ariano}
\ead{dariano@unipv.it}

\ead[url]{www.qubit.it}

\author[dip,infn]{Paolo Perinotti}
\ead{perinotti@fisicavolta.unipv.it}

\address[dip]{QUIT Group, Dipartimento di Fisica ``A. Volta'', via Bassi 6,
  I-27100 Pavia, Italy} 

\address[infn]{Istituto Nazionale di Fisica Nucleare, Sez. di Pavia, via
  Bassi 6, I-27100 Pavia, Italy}

\ead[url]{www.qubit.it}

\begin{abstract}
  We consider the ideal situation in which a space rotation is
  transferred from a quantum spin $j$ to a quantum spin $l\neq j$.
  Quantum-information theoretical considerations lead to the
  conclusion that such operation is possible only for $l\leq j$. For
  $l> j$ the optimal stretching transformation is derived. We show
  that for qubits the present No-Stretching theorem is equivalent to
  the usual No-Cloning theorem.
\end{abstract}


\end{frontmatter}

``No-go'' theorems \cite{940661book} play a crucial role in Quantum
Information Theory \cite{nielchu} and for foundations of Quantum
Mechanics \cite{Clifton:2003p91}.  Among the no-go theorems, the
celebrated No-Cloning
\cite{wootzu,diecks,yuen,Peres:2003p400,Scarani:2005p3025,noteclonref}
is considered the starting point of the field of Quantum Information
itself, lying at the basis of security of quantum cryptography.  Other
relevant no-go theorems are the No-Programming Theorems
\cite{nielchu,fiura,our,our2}, and the No-Universal-NOT
\cite{buzek,demartini}.  The No-Cloning Theorem states the
impossibility of building a machine that produces perfect clones of
the same unknown quantum state.  The No-Programming Theorems state the
impossibility of building a machine that can perform any desired
quantum operation or POVM (positive-operator-valued measure) which is
programmed in a quantum register inside the machine.  Finally, the
No-Universal-NOT states the impossibility of building a device that
reverses a qubit in any unknown quantum state.

The proofs of the No-Cloning and no-programming theorems have a common
feature: in both cases the pertaining ideal transformation should map
pure states to pure states, i.~e. it {\em does not entangle} the
system with the machine.  Therefore, if one supposes that the
transformation is described by a unitary evolution $U$, as dictated by
Quantum Mechanics, the input quantum state $\psi$ is transformed to
$\psi'$ as follows
\begin{equation}\label{Uclon}
  U|\psi\>|\eta\>=|\psi'\>|\eta'(\psi)\>,
\end{equation}
with an auxiliary system (which can be part of the machine, but may
also include the environment) prepared in a {\em reset state} $\eta$
and ending up in a state $\eta'(\psi)$ generally depending on $\psi$.
The argument of the impossibility proof is then to derive a
contradiction by considering the scalar product between different
states at the input and at the output \cite{yuen,yuennote}
\begin{equation}
  \<\phi|\psi\>=\<\phi'|\psi'\>\<\eta'(\phi)|\eta'(\psi)\>,
\end{equation}
and for $|\<\phi'|\psi'\>|< |\<\phi|\psi\>|$, since
$|\<\eta'(\phi)|\eta'(\psi)\>|\leq 1$ one has an overall reduction of
the scalar product, which contradicts the supposed unitarity. In
information theoretical terms a decrease of the scalar product
means an increased state-distinguishability, which would lead to a
violation of the classical data-processing theorem by the machine
regarded as an input-output communication channel.

We will now see that this situation occurs in another no-go
theorem---which we will refer to as {\em No-Stretching
  Theorem}---which {\em forbids stretching a spin while keeping its
  unknown orientation}. In other words, it is impossible to transfer a
spatial rotation from a spin $j$ to a larger spin $l>j$. For a more
general transformation group the situation is more complicated,
because the labels for irreducible representations are usually vectors
rather than (half)integers, and one must find conditions on couples of
such vectors under which transfer from one irrep to another is
impossible.  Increasing dimension of the space carrying the
representation is not a sufficient criterion for impossibility, as one
could easily prove considering the impossibility of covariantly
transforming the representation $U$ for $SU(d)$ to its complex
conjugate $U^*$, which is carried by a space with the same dimension
$d$ \cite{opttransp}. As we will see in the following, it is not just
the angular momentum conservation that matters, since the transfer of
rotation is possible when the spin is decreased to $l<j$.\par

Let us consider a spin $j$ prepared in the coherent state for the
angular momentum $U^{(j)}_g|j,j\>$ with $g$ a generic unknown element
of the group $\SU2$. The state $|j,j\>$ is chosen, with the angular
momentum pointing toward the north pole---however, any other initial
direction would be equivalent. The task is now to transfer the spatial
rotation from the spin $j$ to a different spin $l\neq j$, namely to
get the state $U^{(l)}_g|l,l\>$. If such transfer were physically
feasible there would exist a unitary transformation $W$ such that
\begin{equation}
  W(U^{(j)}_g|j,j\>\otimes|E\>)=U^{(l)}_g|l,l\>\otimes |\theta(g)\>,
\end{equation}
where $|E\>$ is the reset state of the spin-stretching machine, and
$|\theta(g)\>$ is a machine state depending on $g$ [Notice that these
states belong to spaces of different dimensions, since $j\neq l$. For
example one could take $|E\>=|l,l\>\otimes|\omega\>$, $|\omega\>$ the
reset state of an additional ancilla, and then transfer the unitary
rotation $U_g$ from the spin $j$ to the spin $l$.] By taking the
scalar product between vectors rotated with a different $g$, one has
\begin{equation}
  \<j,j|U^{(j)\dag}_hU^{(j)}_g|j,j\>=\<\theta(h)|\theta(g)\>\<l,l|U^{(l)\dag}_hU^{(l)}_g|l,l\>.
\end{equation}
The matrix element of the transformation is just a function of the
second Euler angle $\beta$ of the rotation $h^{-1}g$ (see
Ref.\cite{mess})
\begin{equation}
  \<x,x|U^{(x)\dag}_hU^{(x)}_g|x,x\>=\left(\cos\frac\beta2\right)^{2x},\quad x=j,l,
\end{equation}
whence it is a decreasing function of $x$, since
$0<\left|\cos\frac{\beta}{2}\right|< 1$ (for non-parallel and
non-orthogonal states, i.~e.  $\beta\neq k\pi$, $k$ integer). Then, in
order to preserve the overall scalar product, for $j<l$ we must have
$|\<\theta(h)|\theta(g)\>|>1$, which is impossible, whereas for
decreasing spin $j>l$ we must have $|\<\theta(h)|\theta(g)\>|<1$,
which is allowed by quantum mechanics.

\par
\begin{figure}[h]
\includegraphics[width=6cm]{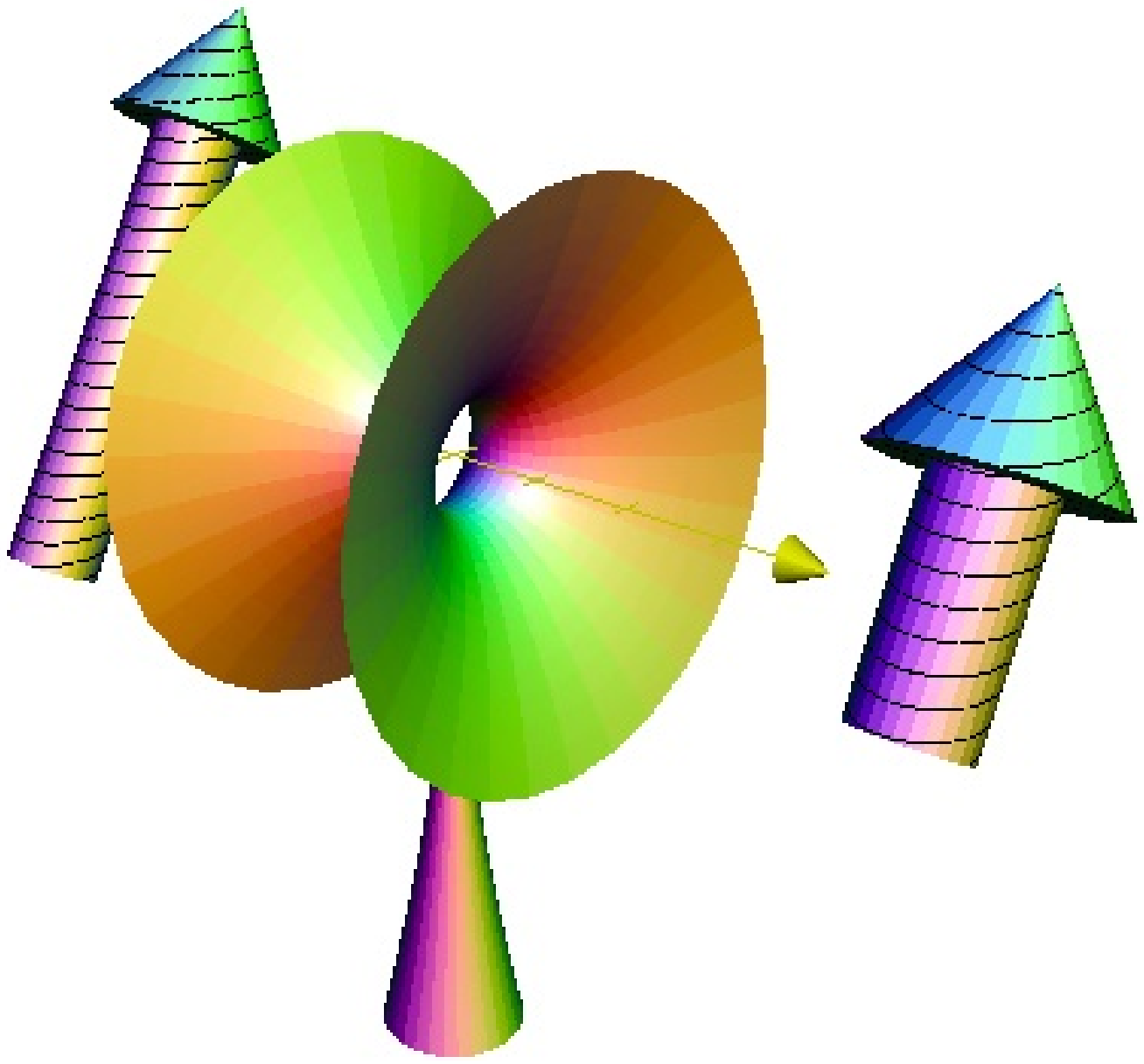}
\includegraphics[width=6cm]{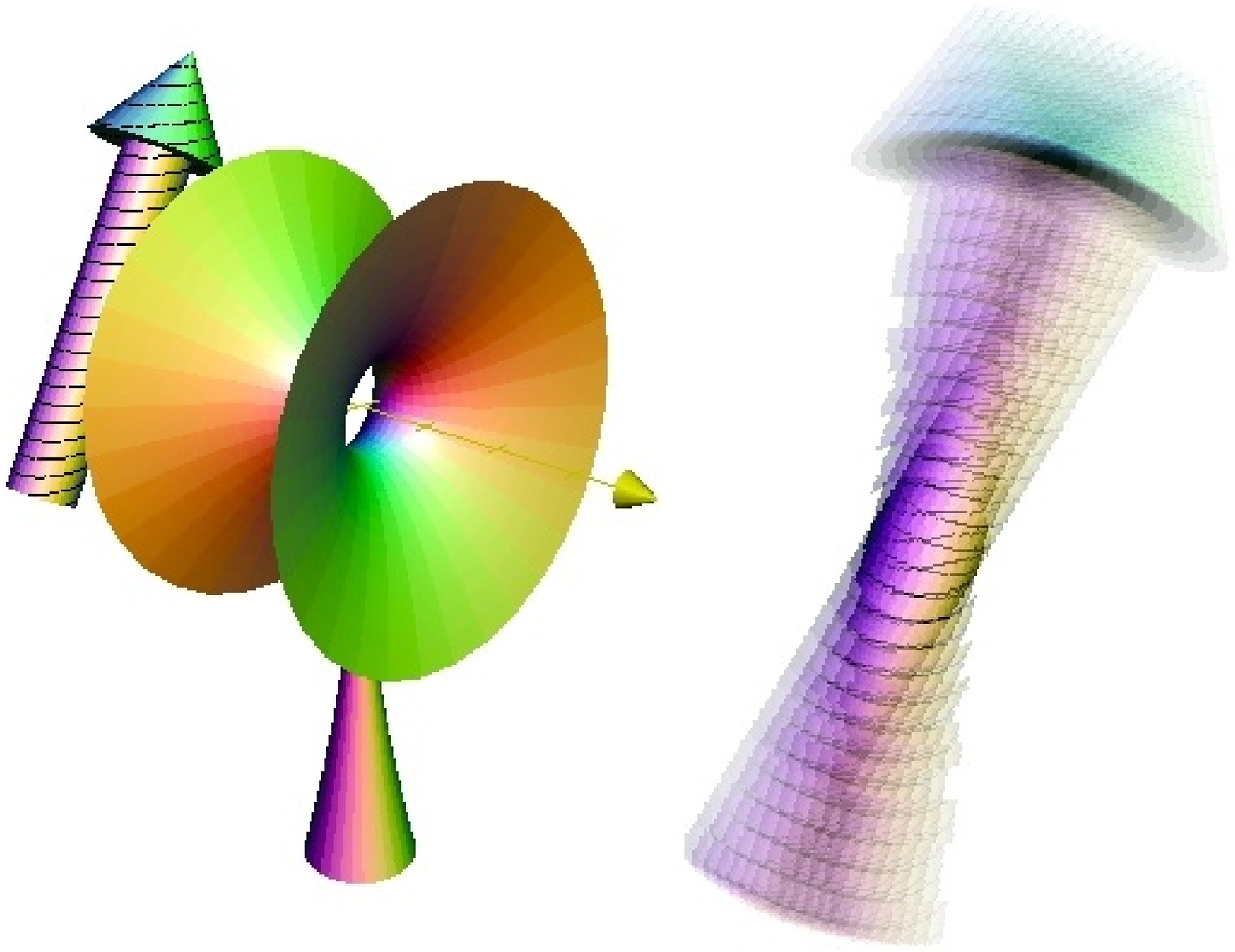}
\caption{\label{pict}\footnotesize The No-Cloning theorem of Quantum
  Mechanics is actually a special case of {\em no-stretching theorem},
  which asserts that unitary transformations cannot be ``amplified''
  to unitaries carrying more information about the parameter of the
  group element, making two nonorthogonal states more
  ``distinguishable''.  For example, there is no machine that takes a
  rotated eigenstate of the $z$-component of the angular momentum and
  produces an output larger angular momentum rotated in the same way.
  In the figure we pictorially represent the no stretching theorem.
  While it is possible to ouput a rotation exactly from a spin $j$ to
  a shorter spin $l<j$ (figure on the top), the same operation cannot
  be achieved exactly when the second spin is larger $l>j$ (figure on
  the bottom). In the latter case the direction is blurred in form of
  a mixing of the output state.}
\end{figure}

 We call the above no-go theorem {\em no-stretching}, since it
forbids to transfer a spatial rotation to a larger spin.  In physical
terms, as can be intuitively understood by figuring a spin as a
vector, this theorem forbids to amplify a signal corresponding to a
spatial rotation by enlarging the vector which is rotated, whereas it
is in principle possible to shorten the vector (as shown in detail in
the following), attenuating the signal (see Fig.  \ref{pict}).\par

If we cannot stretch the spin by keeping the same unknown orientation,
we can anyway try to do our best to keep the orientation by blurring
the state of the spin toward a mixed one. What is then the {\em
  optimal physical stretching map}, which transfers the rotation $g$
from a spin $j$ to a spin $l\neq j$ optimally, e.~g. with the maximum
state-fidelity? For $j<l$ such fidelity must be certainly smaller than
one, whereas for $j>l$ we expect that it can be unit. In technical
terms, in order to be physically achievable the optimal map must be:
1) completely positive (CP)---namely it must preserve positivity also
when applied locally on the system entangled with an ancilla; 2)
trace-preserving; 3) {\em rotation-covariant}, corresponding to the
request that the map transfers the spin rotation. Mathematically, upon
denoting the map as $\rho_l=\map{M}(\rho_j)$ acting on a state
$\rho_j$ of the spin $j$ and resulting in a state $\rho_l$ of the spin
$l$, the covariance of the map is translated to the identity
\begin{equation}
  \map{M}(U^{(j)}_g\rho_j U_g^{(j)\dag})=U_g^{(l)}\map{M}(\rho_j)U^{(l)\dag}_g.
  \label{covar}
\end{equation}
The CP condition is equivalent to the possibility of writing the map
in the Kraus form \cite{kraus}
\begin{equation}
  \map{M}(\rho)=\sum_k M_k\rho M_k^\dag.
\end{equation}
where $M_k$ are linear operators from the input Hilbert space
$\sH_\mathrm{in}$ to the output Hilbert space $\sH_\mathrm{out}$. The
trace-preserving condition corresponds to the constraint
$\sum_kM_k^\dag M_k=I$. Optimality is defined in terms of maximization
of the input-output fidelity
\begin{equation}\label{fidel}
  F:=\<l,l|\map M(|j,j\>\<j,j|)|l,l\>.
\end{equation}

The following Kraus operators $M_k$ give the
optimal map $\map{M}$
\begin{equation}\label{KrausM}
\begin{split}
  M_k=s_{jl}\sum_{m\in I_k}|l,m+k\>\<j,m| \<|j-l|,k;j,-m,l,m+k\>\\
  =s_{jl}\sum_{m\in J_k}|l,m\>\<j,m-k| \<|j-l|,k;j,-m+k,l,m\>,
\end{split}
\end{equation}
where $I_k=[-j,j]\cap[-l-k,l-k]$, $J_k=[-l,l]\cap[-j+k,j+k]$, and
\begin{equation}
  s_{jl}=\sqrt{\frac{2j+1}{2|j-l|+1}}
\end{equation}
where $-|j-l|\leq k\leq |j-l|$, $\<J,M;j,m,l,n\>$ denotes the
Clebsch-Gordan coefficient \cite{mess} for the coupling between the
two spins $j$ and $l$ into their sum $J$.  The Clebsch-Gordan
coefficients in Eq.~\eqref{KrausM} guarantee both trace preservation
and covariance. The above map has been obtained by standard
optimization techniques based on convex analysis. In particular, we
used the Choi-Jamiolkowski representation \cite{choi, Jamiolkowski72}
for CP maps, which exploits the following one-to-one correspondence
between CP maps $\map M$ from $\sH_\mathrm{in}$ to $\sH_\mathrm{out}$
and positive operators $R_{\map M}$ on
$\sH_\mathrm{out}\otimes\sH_\mathrm{in}$
\begin{equation}
  R_{\map M}=\map M\otimes\map I(|\Omega\>\<\Omega|),\quad \map M(\rho)=\Tr_\mathrm{in}[(I_{\sH_\mathrm{out}}\otimes\rho^T)R_{\map M}],
\label{choicorr}
\end{equation}
where $|\Omega\>:=\sum_{j=1}^{d}|\psi_j\>|\psi_j\>$ is a maximally
entangled state, with $d:=\dim(\sH_\mathrm{in})$, the symbol
$\Tr_\mathrm{in/out}$ denotes the partial trace on the Hilbert space
$\sH_\mathrm{in}$ ($\sH_\mathrm{in}$, respectively), and $\rho^T$ is
the transpose of the state $\rho$ on the orthonormal basis
$\{|\psi_j\>\}$. Trace preservation is guaranteed by the condition
$\Tr_\mathrm{out}[R_{\map M}]=I_{\sH_\mathrm{in}}$. The covariance
property Eq.~\eqref{covar} translates to the following commutation
property for $R_{\map M}$ \cite{darlop}
\begin{equation}
  [(U_g^{(l)}\otimes U^{(j)*}_g),R_{\map M}]=0,\quad \forall g\in\SU2.
\end{equation}
Now, it is easy to verify that trace preservation, CP and covariance
properties are all preserved under convex combination of different
maps, which by linearity of Eqs.~\eqref{choicorr} corresponds to
convex combination of Choi-Jamiolkowski operators. Since the fidelity
\eqref{fidel} is linear versus $\map M$ and the set of covariant CP
trace-preserving maps is convex, the optimal map is an extremal point
of such set, namely it cannot be written as a convex combination of
any couple of different maps. Our analysis consists in classifying
extremal points of the set of covariant maps and then looking for the
optimal one.

\begin{figure}[tb]
  \includegraphics[width=9cm]{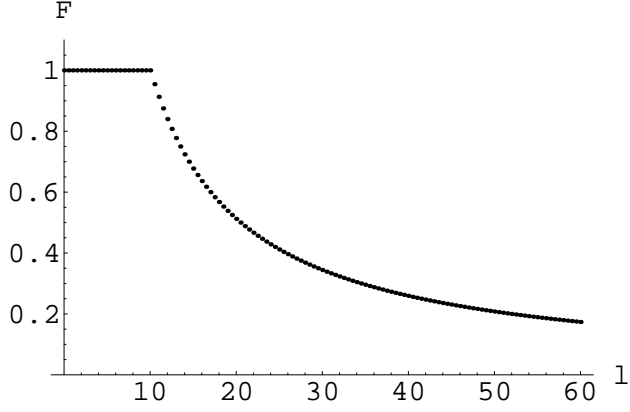}
  \caption{\label{fids}\footnotesize Fidelity of the optimal
    spin-stretching map for $j=10$ as a function of $l$.}
\end{figure}

The derivation of the optimizal map is quite technical, however, it is
easy to check optimality.  Consider the case $j>l$. Then we have
$|j-l|=j-l$.  Applying the map to the state $|j,j\>$ and using
elementary properties of the Clebsch-Gordan coefficients we obtain
\begin{equation}
  \map{M}(|j,j\>\<j,j|)=|l,l\>\<l,l|.
\end{equation}
This proves that the ideal map is exactly achievable for $j>l$. On the
other hand, for $j<l$ we have $|j-l|=l-j$, and the output of the map
applied to $|j,j\>$ in this case is
\begin{equation}
\begin{split}
&\map{M}(|j,j\>\<j,j|)=\\&\frac{2j+1}{2l+1}\sum_{k=j-l}^{l-j}\frac{(2l-2j)!(l+j+k)!}{(2l)!(l-j+k)!}|l,k+j\>\<l,k+j|.
\end{split}
\end{equation}
The fidelity is easily evaluated as
\begin{equation}\label{fidelopt}
  F=\frac{2j+1}{2l+1},
\end{equation}
with plot given in Fig. \ref{fids}.

The optimality of the fidelity (\ref{fidelopt}) can be proved as
follows.  The optimal measurement of the spin direction is described
by the covariant POVM obtained in Ref.  \cite{holevo}
\begin{equation}\label{POVMhol}
  P^{(j)}_g\d g=(2j+1)\d g\, U^{(j)}_g|j,j\>\<j,j|U^{(j)\dag}_g,
\end{equation}
with group integrals normalized as $\int_{\SU2}\d g=1$. This is the
POVM that maximizes the likelihood
\begin{equation}
  L:=\<j,j|P_e^{(j)}|j,j\>,
\end{equation}
of the covariant estimation of $\SU2$ elements on the vector $|j,j\>$
\cite{holevo}, and the maximum likelihood is $2j+1$. Notice that the
POVM in Eq.~\eqref{POVMhol} minimizes all cost functions in the
generalized Holevo class \cite{noi}. We now evolve this POVM with our
map $\map{M}$ with Kraus operators given in Eq.  (\ref{KrausM}).  This
corresponds to apply the dual map $\map{M}^*$ in the reverse order,
i.~e.  from spin $l$ to $j$, corresponding to the Heisenberg picture
(in which we evolve operators instead of states). We thus obtain
\begin{equation}\label{heisenpovm}
  \map{M}^*(P_g^{(l)})=(2l+1)U^{(j)}_g\map{M}^*(|l,l\>\<l,l|)U^{(j)\dag}_g.
\end{equation}
The likelihood of such POVM is
\begin{equation}
\begin{split}
 L= &(2l+1)\<j,j|\map M^*(|l,l\>\<l,l|)|j,j\>=\\
  &(2l+1)\<l,l|\map M(|j,j\>\<j,j|)|l,l\>=(2l+1)F\leq 2j+1,
\end{split}
\end{equation}
and the optimal map saturates this bound.\par

By using the same POVM we can prove that the optimal map preserves the
classical information about the spatial rotation. In order to prove
this statement, let us consider the Kraus operatos in
Eq.~\eqref{KrausM}. Using the identity for the Clebsch-Gordan
coefficients
\begin{equation}
  \<|j-l|,k;j,-m,l,m+k\>=\<|j-l|,-k;l,-m-k,j,m\>,
\end{equation}
and renaming $n=m+k$, the Kraus operators of the dual map $\map
M^*=\sum_kM_k^\dag\cdot M_k$ can be rewritten as follows
\begin{equation}
\begin{split}
  M_k^\dag=&s_{jl}\sum_{m\in I_k}|j,m\>\<l,k+m| \<|j-l|,k;j,-m,l,k+m\>\\
  =&s_{jl}\sum_{n\in J_k}|j,n-k\>\<l,n| \<|j-l|,k;j,k-n,l,n\>\\
  =&s_{jl}\sum_{n\in J_k}|j,n-k\>\<l,n| \<|j-l|,-k;l,-n,j,-k+n\>.
\end{split}
\end{equation}
Considering that $s_{lj}=\sqrt{\frac{2l+1}{2j+1}}s_{jl}$, it is now
immediate to notice that the dual map $\map M^*$ for the case $l>j$
coincides with the direct map for input spin $l$ and output $j$, apart
from a multiplicative constant $\frac{2j+1}{2l+1}$, since the Kraus
operator $M_k^\dag$ of $\map M^*$ coincides with the Kraus operator
$M_{-k}$ of $\map M$ from $l$ to $j$. Then,
\begin{equation}
  \map
  (2l+1)\map M^*(|l,l\>\<l,l|)=(2j+1)|j,j\>\<j,j|.
\end{equation}
This implies that the conditional probability distribution
\begin{equation}
  p(g|h)=\Tr[P^{(l)}_g\map M(U^{(j)}_h|j,j\>\<j,j|U_h^{(j)\dag})],
\end{equation}
for the outcomes of the measurement described by the POVM $P^{(l)}_g$
at the output of the optimal stretching channel is exactly the same as
that of $P^{(j)}_g$ at the input
\begin{equation}
  q(g|h)=\Tr[P^{(j)}_g U^{(j)}_h|j,j\>\<j,j|U_h^{(j)\dag}].
\end{equation}
Since the mutual information of the two random variables $g,h$ is a
functional of the conditional probability, $p(g|h)=q(g|h)$, this
implies that the mutual information obtained by the POVM $P_g^{(j)}$
at the input is preserved at the output. Therefore, the optimal
spin-stretching map preserves the mutual information.\par

For {\em qubits} the No-Cloning theorem is equivalent to the
No-Stretching theorem. Indeed, perfect cloning from $m$ to $n>m$
copies is equivalent to stretching the total angular momentum from
$j=\frac{m}{2}$ to $l=\frac{n}{2}$. Moreover, the optimal fidelity for
$m\to n$ cloning is given by \cite{werner98}
\begin{equation}
F=\frac{m+1}{n+1}, 
\end{equation}
which coincides with Eq. \ref{fidelopt}. Clearly, No-Cloning for
qubits implies No-Cloning for qudits. For qtrits or generally larger
dimension $d>2$, what the No-Stretching theorems forbid is to transfer
a group transformation covariantly to a system carrying more
information about such transformation. However, this condition is
harder to state in precise mathematical terms involving parameters of
irreducible input and output representations.

In conclusion, we have seen that it is forbidden to stretch a spin
while keeping its unknown orientation, a new no-go theorem which we
call {\em No-Stretching Theorem}.  We have seen that this is not due
to conservation laws, since the transformation in the opposite
direction---i.~e. decreasing the angular momentum---is possible
perfectly (this is non obvious). The No-Cloning theorem is a special
case of the no-stretching Theorem, and for qubits the optimal
spin-stretching $j\to l$ transformation coincides with the optimal
cloning from $m=2j$ to $n=2l$ copies.\par

This work has been supported by the EC through the project CORNER.

\end{document}